# Macroscopic non-equilibrium thermodynamics in dynamic calorimetry

**J.-L. Garden***

*Centre de Recherches sur les Très Basses Températures, CNRS, laboratoire associé à l'Université Joseph Fourier et à l'Institut National Polytechnique de Grenoble, BP 166, 38042 Grenoble Cedex 9, France.*

## Abstract

What is really measured in dynamic calorimetric experiments is still an open question. This paper is devoted to this question, which can be usefully envisaged by means of macroscopic non-equilibrium thermodynamics. From the pioneer work of De Donder on chemical reactions and with other authors along the 20$^{th}$ century, the question is tackled under an historical point of view. A special attention is paid about the notions of frequency dependent complex heat capacity and entropy production due to irreversible processes occurring during an experiment. This phenomenological approach based on thermodynamics, not widely spread in the literature of calorimetry, could open significant perspectives on the study of macro-systems undergoing physico-chemical transformations probed by dynamic calorimetry.

Keywords: ac-calorimetry; Non-equilibrium thermodynamics; Affinity; Entropy production; Frequency dependent heat capacity; Complex heat capacity.

* Corresponding author. Fax: 33 (0) 4 76 87 50 60

*E-mail address:* jean-luc.garden@grenoble.cnrs.fr (J.-L. Garden)

## 1. Introduction

In calorimetric heat capacity measurements, a sample is perturbed by an input thermal power and the resulting temperature variation is measured. Also, a temperature program can be predetermined and the resulting heat flow can be measured. Under specific experimental conditions, the heat capacity of the sample is defined by the ratio of this thermal power on the measured temperature rate. Nevertheless, when the heat flow is supplied on a time scale smaller than the internal reorganization time of the sample degrees of freedom, the measured heat capacity is the result of a non-equilibrium experiment. What is then exactly measured by the experimentalist? The result of the measurement is sometimes called the apparent heat capacity. The same type of question can be asked in modulated calorimetric experiments when the input heat flow frequency is higher than the frequency of the degrees of freedom constituting the heat capacity of the sample. This yields the so-called frequency dependant complex heat capacity with a real and an imaginary component. Nowadays, no clear consensus exists on what is really measured during theses dynamic calorimetric measurements.

This paper aims at demonstrate that the formalism of macroscopic non-equilibrium thermodynamics can be very helpful to envisage these questions. We propose to aboard the problem on a historical point of view. For example, we will see that the notions of non-equilibrium heat capacity and frequency dependent complex heat capacity have been already envisaged for a long time. After the present introduction, in the section 2, the historical background of the frequency dependent complex heat capacity in calorimetry is given. The definitions of macroscopic non-equilibrium thermodynamics and dynamic calorimetry are then also provided. In the section 3, dynamic calorimetric experiments are envisaged on a qualitative manner. The working assumptions of what we consider as an ideal dynamic calorimetric experiment are previously given. Then, the link existing between calorimetry, non-equilibrium thermodynamics and kinetics is envisaged qualitatively. In the section 4 and 5, dynamic calorimetric experiments and complex heat capacity notion are tackled on a quantitative manner through the works of different authors. All along the paper, a special attention is paid about the notion of rate of production of entropy generated during non-equilibrium physico-chemical transformations. Although we have focused on modulated calorimetric experiments and complex heat capacity, in the last section (section 6) we treat an example of dynamic calorimetric experiments, the dynamic differential scanning calorimetry (DSC), for which macroscopic non-equilibrium thermodynamics can be also applied. Then, we show that the formula generally found in the literature of the averaged entropy production over one period of temperature oscillation is simply issued from a peculiar case of irreversible process. Next, before the conclusion, we provide our point of view on the physical meaning of the imaginary part of the complex heat capacity.

## 2. Historical background, definitions and assumptions

### *2.1. Historical background of frequency dependent complex heat capacity in calorimetry.*

At the beginning of the 20$^{th}$ century Corbino stated the basis of modulated calorimetric experiments [1, 2]. At the end of the sixties, Sullivan and Seidel improved the technique with the so-called steady-state ac-calorimetry method, useful in low temperature specific heat measurements [3]. Heat capacities were anyway already measured by Kraftmakher and others from modulated temperature experiments. An interesting review on the subject has been written by Kraftmakher [4].

Few years later in 1970, Gobrecht and co-workers had the original idea to replace the linear temperature ramp of usual differential scanning calorimeters (DSC) by a modulated one [5]. It was the birth of temperature modulated differential scanning calorimetry (TMDSC). This article contained all the concepts used today in modern TMDSC measurements: use of a complex heat capacity; separation of the vibrational and configurational modes of the heat capacity; application to the glass transition; cole-cole plot of the complex heat capacity.

At the beginning of the nineties, Reading and co-workers refund the principle of the TMDSC [6]. With the use of a deconvolution, Reading proposed to separate the calorimetric signal into a reversing and a non-reversing component. Next, Schawe proposed a new physical interpretation of the two components measured in TMDSC. Since, a very famous and interesting dispute has opposed the two authors [7-9]. The interpretation of Schawe results in a new separation of the TMDSC signal into two heat capacity components. One is called the storage heat capacity, and the other one the loss heat capacity. From this point of view, the heat capacity measured in TMDSC experiment is a complex heat capacity with a real and an imaginary part. This usual equilibrium thermodynamic quantity must thus be regarded as a generalized dynamic susceptibility such as non-equilibrium response derived from dielectric or magnetic susceptibility measurements:

$$C^* = C' - iC'' = C_\infty + \frac{C_0 - C_\infty}{1 + i\omega\tau} \quad (1)$$

where

$$C' = C_\infty + \frac{C_0 - C_\infty}{1 + (\omega\tau)^2} \quad (2)$$

is the storage frequency dependent heat capacity, and:

$$C''=\frac{(C_0-C_\infty)\omega\tau}{1+(\omega\tau)^2} \quad (3)$$

is the loss frequency dependent heat capacity. *C'* and *C''* satisfy the so-called Kramers-Kronig dispersion relations. $\omega$ is the angular frequency of the modulated temperature. $C_\infty$ is the heat capacity related to the infinitely fast degrees of freedom of the system as compared to the frequency (generally vibrational modes or phonons bath), and $C_0$ is the total contribution at equilibrium (the frequency is set to zero) of the degrees of freedom, fast and slow, of the sample. The time constant $\tau$ is the kinetic relaxation time constant of the non-equilibrium degree of freedom.

These three last formulas have been already derived a long time ago with the formalism of the linear response theory, especially from the work of Birge and Nagel who measured the frequency dependent heat capacity of liquids with the 3ω method [10-14]. At the same time, frequency dependent complex heat capacity was also envisaged by Christensen [15]. Specific heat spectroscopy with the 3ω-method was also tackled in the following references [16-19]. An other original approach is due to Toda and co-workers, who derived the complex heat capacity directly through a pure kinetic approach during melting of polymer crystals [20, 21]. Generally, the formula of the complex heat capacity was always derived from the linear response theory. In this approach, the heat capacity is seen as the linear response of a small perturbation of the entropy of the system (or enthalpy), knowing that the measured temperature is the conjugated thermodynamic variable of the entropy. Derivations of the complex heat capacity by the linear response theory can be found for example in the following references [22-31]. However, this approach is outside the scope of this article.

Principally since the birth of the TMDSC method, lot of scientists have tried to better understand the notion of complex heat capacity and particularly the meaning of the imaginary part of *C** without any clear conclusions. For commenting our assertion, let cite some typical sentences, which can be frequently encountered in the literature:

- In 1997, Simon and McKenna wrote an interesting review on frequency dependent complex heat capacity and one of their conclusions is [32]: "*…the subsequent discussion demonstrated that there is no consensus concerning the interpretation of dynamic heat spectroscopy measurements*".

- In 1997, in a special issue of Thermochimica Acta on TMDSC, Höhne wrote a letter called "*Remark on the interpretation of the imaginary part C'' of the complex heat capacity*" [33]. He proposed to use thermodynamics of irreversible processes to envisage the question.

- In 1998, in a special issue of Journal of Thermal Analysis and Calorimetry on TMDSC, Scherrenberg and co-workers wrote: "*The physical meaning of the imaginary heat capacity in regime Ib is still subject to debate*" [29]. Farther, in the same journal, Buehler and co-workers wrote: "*Basically, there is no well-founded physical or thermodynamical interpretation of* $C_p''$" [34].

- In 2000, Buehler and Seferis wrote in the abstract of their paper: "*It also explored the influence of sample thickness on heat flow phase, without using a complex heat capacity of doubtful physical meaning*". And farther, after a discussion about their own interpretation, they gave a detailed table taking into account different interpretations of the imaginary part of the heat capacity according to various authors [35].

- In 2001, Simon wrote: "*The frequency dependence of the specific heat in an equilibrium (ergodic) system has been variously related to fluctuations in enthalpy, in temperature, and in entropy, although general agreement has not been reached*" [36].

- In 2002, Jiang and co-workers wrote: "*The problem with this approach is that, at present, there is no universally acknowledged interpretation of the meaning of the out-of-phase component $C_p$'*'" [37].

- Also in 2002, in an interesting published email exchange, Schick and Saruyama reported that the "*Frequency dependence of heat capacity and its interpretation is one of the still open questions in calorimetry*" [38].

A good review of these remaining questions is given by Claudy in his recent book [39]. To our point of view, the real part of the complex heat capacity with its frequency dependency can be clearly understood, but the physical meaning of the imaginary part or the loss part of the heat capacity remains confuse. In usual dynamic susceptibility measurements, imaginary parts of generalized susceptibilities are well physically understood and always linked to heat dissipation inside the sample. For calorimetric measurements when the perturbing parameter is already heat, what does heat dissipation mean? Does it have even a physical sense? If the imaginary part of the complex heat capacity is linked to thermal dissipation, where is passed the heat (heat lost) during one period of the temperature oscillation, knowing that the experiment can be realized in an adiabatic manner (no heat has time to release towards the heat bath over one period)? An attempt of the response of these last questions is envisaged in the last section of this present manuscript.

### *2.2. Definition of dynamic calorimetry.*

Calorimetry is an experimental technique concerned by measurements of amounts of heat exchanged by a sample with its surrounding. Sometimes, theses quantities of heat are produced (or absorbed) by the sample itself when a physico-chemical transformation occurs (enthalpy measurements due to the variation of an external parameter such as the pressure, the temperature, the magnetic field, the adding of a constituent, etc…). Sometimes, the experimentalist itself provides (or released) heat to the sample for probing its structure or its internal degrees of freedom (heat capacity measurements). In all case, these measurements are realized with a thermometer. At the scale of the sample, composed of a very large number of sub-systems, heat and temperature are macroscopic thermodynamic variables. They result on the average taken over all the sub-systems strongly linked together, which constitute an entire thermodynamic macroscopic system. $T$ the temperature and $Q$ the heat are quantities of great importance in the field of thermodynamics. For example, the ratio of the heat exchanged between the sample and its surroundings to the absolute sample temperature is the external entropy variation of the system. Also, the ratio of this quantity of heat to the temperature variation of the system is the heat capacity of the system. If the temperature, its variation and the heat capacity are measured, then the enthalpy, the entropy and the Gibbs free energy variations can be derived. The only experimental method which permits a direct access to these thermodynamic quantities is calorimetry. Hence, it is obvious to state that this experimental method is intimately connected to the theoretical approach of equilibrium and non-equilibrium thermodynamics. Dynamic calorimetry can have two different significations:

- dynamic in the sense of a variation of the sample temperature.
- dynamic in the sense that the measured quantities are not in thermodynamic equilibrium and consequently can not be considered as static quantities.

We will see that these two definitions are dependant. In this paper, we adopt the last definition. It is the same one adopted by Birge and Nagel [10-14], and by Jeong (see the review on dynamic calorimetry [28]). Indeed, macroscopic thermodynamics is concerned by time average of variables, which are in equilibrium and considered as static. Non-equilibrium thermodynamics is concerned by dynamic variables, which are not in thermodynamic equilibrium. When kinetic relaxation times of thermal events under study become long compared to the time scale of the measurement, thermodynamic variables have no time to reach their equilibrium values. What is then really measured by the experimentalist? What are the conditions for a calorimetric experiment to be considered as static or dynamic? We will see along this paper that these questions are also related to frequency dependent heat capacity measurements when the period of the oscillating temperature becomes smaller than the kinetic relaxation time of thermal events under study.

### *2.3. Definition of macroscopic non-equilibrium thermodynamics.*

We consider only classical finite macroscopic system with macroscopic thermodynamic variables such as volume, temperature, pressure and others, which are subdued only by the first and the second laws of thermodynamics. Microscopic thermodynamics and statistical mechanics governed by probabilities and fluctuations are not used. Thermodynamic systems (in fact sample under calorimetric study) are uniform regarding to the intensive variables such as the pressure, the temperature, but are in a non-equilibrium state regarding to peculiar internal degrees of freedom. Before entering in the connections existing between calorimetry and thermodynamic irreversibility, let us give a brief historical summary of macroscopic non-equilibrium thermodynamics.

### *2.4. Historical survey of macroscopic non-equilibrium thermodynamics*

For a good historical description of non-equilibrium thermodynamics, see references [40-43]. At the end of the 19th century, Gibbs defined the basis of classical equilibrium thermodynamics [44]. After his work, which is still extensively used nowadays, the first approach envisaging the field of non-equilibrium thermodynamics on a general manner is due to Onsager in 1931 [45]. From the principle of microscopic reversibility, Onsager establishes the so-called reciprocal relations. This work gives for the first time a clear formal explanation of irreversible processes such as Fourier's law, Thomson's effect and others. From this approach, scientists have discussed the connection existing between macroscopic and microscopic thermodynamics. All the important theorems issued from this period are based on the fundamentals laws of statistical mechanics. Most of them are based on an important assumption: at microscopic level, fluctuations occurring near equilibrium have the same decreasing exponential behavior towards equilibrium that macroscopic thermodynamic variables, which have been moved aside equilibrium by an external force. For a good survey of the subject, see the non-exhaustive following references [46-56]. In the following, we will see that the important notions of generalized thermodynamic forces and associated responses (generalized thermodynamic fluxes) are of great interest in dynamic calorimetry.

On the other hand, in the twenties at the Université libre de Bruxelles, De Donder defined the thermodynamic state function, $A$, the affinity, which represents the driving force of a chemical reaction [57, 58]. Chemical reactions are always non-equilibrium processes. At thermodynamic equilibrium, no reaction occurs and the affinity is equal to zero. The concept of affinity has been next generalized by different authors, such as Prigogine, Defay, De Groot and Mazur [41, 59]. Nowadays, generalized affinities are used on a very general manner to represent driving forces of any irreversible thermodynamic processes. See, for an interesting use of De Donder's thermodynamics and generalized affinities applied to internal reorganizations with relaxation phenomena, the work of Cunat [60]. Even for phase transitions or phase transformations, the affinity (difference of chemical potential between the phases) is the thermodynamic force driving the advancement of the transitions. In the section 4, through the works of De Donder, Prigogine and Defay, we will envisage quantitatively how the concept of generalized affinity can play a key role in dynamic calorimetric experiment. Before that, let us provide the frame of macroscopic non-equilibrium thermodynamics applied to calorimetry, and envisage qualitatively the definition of an irreversible calorimetric experiment.

## 3. Qualitative approach of irreversible thermodynamics in calorimetric experiments

### *3.1. Working Assumptions for calorimetric experiments*

Let be a sample with a heat capacity $C$ at a temperature $T$ linked by a heat exchange coefficient $K$ to a thermal bath with a constant temperature $T_0$ (see Fig. 1). The sample (with its addenda) represents a macroscopic thermodynamic system.

We assume that this thermodynamic system is represented by three independent state variables $(p,T,\xi)$. $p$ the pressure and $T$ the temperature are the two physical variables, and $\xi$ is the chemical variable (the definition of this variable is given in the section *4.1.1.*). More generally this variable can represent a generalized order parameter connected to a specific internal degree of freedom of the sample. If the pressure $p$ is maintained constant during the entire calorimetric experiment, then the state of the system is defined by the set $(T,\xi)$. More precisely, the evolution of the state of the system is given by the two functions $T(t)$ and $\xi(t)$. The thermodynamic transformation is represented by a curve in the diagram $\{T,\xi\}$ (see Fig. 2).

Let consider that the system is a thermodynamic closed system. This is to say that the system can only exchange energy with the outside world (no exchange of matter).

The system is also considered in thermal equilibrium. This is to say that there is no temperature gradient inside the system, or the temperature is homogeneous in all parts of the system at any time. In general, this condition can be fulfilled because the function $T(t)$ is known and controlled by the calorimetrist. For a given temperature variation $\Delta T$, realized in the time interval $\Delta t$, if the internal heat relaxation time is less than $\Delta t$, then the sample is homogeneous in temperature. The internal heat relaxation time of the system is linked to the thermal diffusivity of the sample and its thickness. Hence, this condition is reached if the volume of the sample is small for a given value of the thermal diffusivity of the sample.

Finally, let consider that the system is in mechanical equilibrium, which means that the pressure is homogeneous and there is no fast volume variation inside the sample during the experiment. This assumption is true if the pressure of the system is kept constant during the experiment.

From these last conditions, $\xi$ is the only variable sensitive to a non-equilibrium situation. This is to say, for a given variation of the state variable $T$, it is possible that $\xi$ does not reach its equilibrium value. Hence, in an irreversible calorimetric experiment the set $(T,\xi)$ does not represent a state of equilibrium.

*3.2. Reversible and irreversible calorimetric experiments*

De Donder and the members of his school wrote the second law of thermodynamics as the following:

$\delta Q' = TdS - \delta Q \geq 0$ (4)

where the letter $\delta$ takes into account that heat is not a state function and not an exact total differential. $\delta Q'$ is the uncompensated heat of Clausius. This is the quantity of heat produced within the system when an irreversible process occurs. $T$ is the absolute temperature, $\delta Q$ the quantity of heat exchanged by the system with the outside world and $dS$ the infinitesimal total entropy variation. First of all, let us consider a thermodynamic system at an initial equilibrium state A. One assumes that the system undergoes a physico-chemical transformation. If the transformation drives the system from an equilibrium state A to an other equilibrium state B, and if it is an equilibrium transformation (a transformation which proceeds by a succession of equilibrium states) then the amount of heat exchanged between the system and its surroundings is:

$Q_1 = T\Delta S_1$ (reversible transformation) (5)

If the transformation occurs outside equilibrium, then on the basis of De Donder's definition of the second law of thermodynamics:

$Q_2 = T\Delta S_2 - Q'$ (irreversible transformation) (6)

The entropy $S$ being a state function, if A and B are the same in the two experiments, then $\Delta S_1 = \Delta S_2$ and consequently:

$Q' = Q_1 - Q_2$ (7)

Therefore, $Q'$ is the difference between the amount of heat exchanged by the system with its surroundings, for a reversible and an irreversible transformations respectively which drive the system from the same equilibrium state A to the same equilibrium state B.

In calorimetric experiments, the situation is slightly different because the quantity of heat supplied to (or released from) the system by the outside world is controlled by the experimentalist. In this case, (5) and (6) are written:

$Q = T\Delta S_1$ (8)

for a reversible experiment, and:

$Q = T\Delta S_2 - Q'$ (9)

$Q' > 0$ for an irreversible experiment. $Q$ is supposed to be the same in the two experiments. It implies that $\Delta S_2 \neq \Delta S_1$. Therefore, after $\Delta t$, if B is in a state of equilibrium in the reversible experiment, it is not the case in the irreversible experiment, because $S_{B2} \neq S_{B1}$ ($S$ being a state function). This is the reason why thermodynamics of irreversible processes is called non-equilibrium thermodynamics. In this "Gedanken experiment", since we compare two calorimetric experiments with the same amount of heat supplied to the same sample, thus inevitably $\Delta t$ must be shorter in the irreversible experiment than in the reversible experiment. This is to say that the heat flow supplied to (or released from) the sample is higher in the irreversible experiment than in the reversible one. In the irreversible case, relaxation phenomena inside the sample (kinetics) cannot be neglected. In this case, temperature rates of the sample are high. Hence, we see the link between fast temperature ramp and dynamic calorimetry. Let insist that the irreversibility of a calorimetric experiment is not an absolute notion. It depends only on the time interval (time scale of the measurement) over which the quantity of heat $Q$ is supplied to the sample. In the irreversible experiment, the uncompensated heat of Clausius is produced in a time interval that lasts longer than this characteristic time scale, $\Delta t$, of the measurement. The fact that a process could be considered reversible or irreversible depending on the time scale of the observation is well expressed by Chandrasekhar in the reference [61]: "*Quite generally, we may conclude with Smoluchowski that a process appears irreversible (or reversible) according as whether the initial state is characterized by a long (or short) average time of recurrence compared to the times during which the system is under observation.*"

### *3.3. Entropy production*

The second law of thermodynamics (4) can be rewritten in a different way:

$dS = d_eS + d_iS$ (10)

where $dS$ is the infinitesimal entropy variation of the system, $d_eS$ is the infinitesimal entropy variation exchanged between the system and the surroundings, $d_iS$ is the infinitesimal entropy produced by irreversible processes occurring within the sample. More explicitly:

$d_eS = \delta Q/T$ (11) and $d_iS = \delta Q'/T \geq 0$ (12)

During a calorimetric experiment, the quantity of heat $Q$ is exchanged between the experimentalist and the sample during the finite time interval $\Delta t$. Thus the instantaneous heat flow exchanged between the system and its surroundings over the time interval $\Delta t$ is written:

$$P = \delta Q/dt = Td_eS/dt = TdS/dt - Td_iS/dt \quad (13)$$

The time derivative term $d_iS/dt$ is the so-called rate of the production of entropy (or simply entropy production). Knowing that this term (equal to zero only when the experiment is realized in a reversible manner) is linked to a real positive quantity of heat produced within the sample, it is legitimate to ask for the following questions: is this positive quantity of heat perturbing the heat capacity measurement? Is it the cause of the frequency dependent heat capacity in modulated calorimetry measurements? In the next section, we will see that it is not the case.

## 4. Quantitative approach of macroscopic non-equilibrium thermodynamics in calorimetric experiments.

This section is principally based on the work of De Donder, Prigogine and Defay. From 1927 to 1934, De Donder has regrouped his works on chemical irreversible processes in three books [57, 58]. Important formulas of the heat capacity were derived from the two principles of thermodynamics. The first formula of non-equilibrium heat capacity measured at constant affinity was derived by De Donder. Later, in 1946 and 1950, Prigogine and Defay pushed further the reasoning of De Donder, and derived for the first time the general formula of the measured heat capacity during non-equilibrium transformations [59, 62]. These formulas, not very known, can be of great interest in dynamic calorimetry, and we propose to provide the details of their derivations in this section.

### *4.1. The thermodynamics of Théophile De Donder*

#### *4.1.1. Affinity and degree of advance of a reaction*

De Donder was the first to generalize the classical Gibbs's equilibrium thermodynamics to irreversible processes occurring during chemical reactions. Among his monumental work, one of the most important discoveries was to find the quantitative expression of the driving force of chemical reactions. De Donder expressed this force by a new thermodynamic state function $A$, the affinity. It can be regarded as the cause of the advance of chemical reactions. Let consider a simple chemical reaction:

$$\nu_a\mathrm{A} + \nu_b\mathrm{B} \rightarrow \nu_c\mathrm{C} + \nu_d\mathrm{D} \quad (14)$$

where A, B and C, D, are the reactants and the products respectively and $\nu_a$, $\nu_b$, $\nu_c$, $\nu_d$ are the stochiometric coefficients. The affinity is defined by De Donder as follows:

$$A = (\nu_a\mu_a + \nu_b\mu_b) - (\nu_c\mu_c + \nu_d\mu_d) \quad (15)$$

where $\mu_x$ is the chemical potential of the constituent $x$. The genius idea of De Donder was to express the uncompensated heat of Clausius as a product of a generalized thermodynamic force

(the affinity) with a generalized thermodynamic flux (the variation of the degree of advance of the reaction):

$\delta Q' = Ad\xi \geq 0$ (16)

The degree of advance of the reaction is a thermodynamic state variable, which represents the advancement of a chemical reaction. It is defined by De Donder as follows:

$N_x(t) = N_x(0) + \nu_x\xi$ (17) or $d\xi = dN_x/\nu_x$ (18)

with $\xi = 0$ at the initial state, and where $N_x$ is the number of mole of the constituent $x$. The initial mole number of each constituent being known, the chemical reaction is entirely defined by the degree of advance of the reaction. This variable of state can also characterize a phase transformation. Hence, it represents the changes between the different constituents during a chemical reaction or the changes between the different phases of a system during a phase transformation. With these definitions, the irreversible positive entropy produced during physico-chemical reaction takes a simple expression as the product of the chemical force with the induced flux (rate of reaction, $v$):

$$\sigma_i = \frac{d_iS}{dt} = \frac{1}{T}\frac{dQ'}{dt} = \frac{A}{T}\frac{d\xi}{dt} = \frac{A}{T}v \geq 0 \quad (19)$$

The rate of the uncompensated heat of Clausius, or the thermal power of irreversibility is simply given by:

$P_i = T\sigma_i$ (20)

As the affinity is the force driving the system towards equilibrium when it is moved aside equilibrium, it is also possible to derive the following fundamental equation [57, 58]:

$A = -\,\partial G/\partial\xi)_T = T\partial S/\partial\xi)_T - \partial H/\partial\xi)_T$ (21)

where $G$ is the Gibbs's free energy, representing the chemical potential in the set variable $(T, \xi)$ ($p$ being constant). $\partial H/\partial\xi)_T$ is the heat of reaction at constant pressure and temperature, and $\partial S/\partial\xi)_T$ is the entropy variation due to the reaction at constant pressure and temperature.

At equilibrium the affinity and the rate of reaction vanish together and we have:

$\partial H/\partial\xi)_T^{eq} = T\partial S/\partial\xi)_T^{eq}$ (22)

*4.1.2. Total differential of the affinity*

It was rigorously demonstrated that the affinity is a state function [63, 64]. Thus, De Donder has differentiated this state variable with respect to the other independent variables of the system:

$dA = \partial A/\partial T)_\xi dT + \partial A/\partial\xi)_T d\xi$ (23)

which for simplicity is rewritten as the following:

$dA = \alpha dT - \beta d\xi$ (24)

With (21) we have:

$\beta = \partial^2 G/\partial\xi^2)_T$ (25)

The coefficient $\beta$ (the second derivative of the Gibbs's free energy) is always positive around equilibrium because $G$ is minimum at equilibrium. This results on the stability of the equilibrium state after a perturbation [65, 66]. From reference [58], there is also:

$$\alpha = \partial A / \partial T)_\xi = \partial S / \partial \xi)_T = \frac{\partial H / \partial \xi)_T + A}{T} \quad (26)$$

This equation is called the Berthelot-De Donder's formula. It can be derived directly from the definition (21) of the affinity. Physically, it means that, at constant temperature and pressure, the affinity is a thermodynamic potential for the system undergoing a physico-chemical transformation, a driving force, which vanishes when the system is at equilibrium.

*4.1.3. Heat capacity at constant affinity*

The first law of thermodynamics states that, at constant pressure, the amount of heat exchanged between the system and the outside world is equal to the variation of the enthalpy of the system:

$\delta Q = dH$ (27)

In this case, an important formula of the heat capacity of the system in the set of thermodynamic variables *(T, ξ)* can be derived [58, 59]:

$$C_{mes} = \frac{\delta Q}{dT} = C_{\xi} + \partial H / \partial \xi)_T \frac{d\xi}{dT} \quad (28)$$

where $C_{\xi} = \partial H/\partial T)_{\xi}$ is the heat capacity at constant composition of the system, often called the true heat capacity of the system. This basic thermodynamic formula was written by De Donder in another way in order to make more evident the role of the affinity. With (24) and (28) making $dA = 0$, and with (25) and (26), he derived the heat capacity at constant affinity (see reference [58] page 58):

$$C_{mes} = C_{A_0} = C_{\xi} + \partial H / \partial \xi)_T \frac{[\partial H / \partial \xi)_T + A_0]}{T\partial^2 G / \partial \xi^2)_T} \quad (29)$$

which is in fact the first time that a quantitative expression of the heat capacity during non-equilibrium event is derived. At equilibrium, the affinity vanishes and the expression of the heat capacity is:

$$C_{mes} = C_{\xi} + \frac{[\partial H / \partial \xi)_T^{eq}]^2}{T\partial^2 G / \partial \xi^2)_T^{eq}} \quad (30)$$

In this paper we consider:

$$\eta_{eq} = \frac{[\partial H / \partial \xi)_T^{eq}]^2}{T\partial^2 G / \partial \xi^2)_T^{eq}} \quad (31)$$

Hence, the heat capacity measured during a reversible calorimetric experiment can be simply written as the following:

$C_{mes} = C_{rev} = C_{\xi} + \eta_{eq}$ (32)

In accordance with the stability condition of the state of equilibrium ($\beta_{eq} > 0$), $C_{mes}$ is always greater than the true heat capacity $C_{\xi}$ of the sample during a reversible experiment. Let us note that this conclusion is true even if the physico-chemical transformation is exothermic or endothermic during the experiment.

*4.1.4. Entropy production*

At this level, we would like to insist on an important point that is never mentioned in lots of publications on the subject. It should be pointed out that the term of entropy production is hidden in the general formula (28) of the heat capacity, which can be easily rewritten:

$$C_{mes} = C_{\xi} + T\partial S / \partial \xi)_T \frac{d\xi}{dT} - T\frac{d_iS}{dT} = C_{\xi} + T\partial S / \partial \xi)_T \frac{d\xi}{dT} - \frac{\delta Q'}{dT} \quad (33)$$

A simple and naïve interpretation of the equation (33) could be as follows: the term of uncompensated heat of Clausius by unit of temperature (or “uncompensated heat capacity of Clausius”) may be subtracted to the equilibrium heat capacity and may be consequently responsible of the decrease of the measured heat capacity in non-equilibrium calorimetric experiments. Knowing that generally the non-equilibrium measured heat capacity is smaller than the heat capacity measured at equilibrium, this term, equal to zero only at equilibrium, could be responsible of the frequency dependent heat capacity effects during temperature modulated calorimetric experiment. Unfortunately, the situation is not so simple, and we will see later that this term is neglected in the derivation of the frequency dependent complex heat capacity by means of the De Donder's formalism.

*4.2. Generalized affinities*

Nowadays, it is well known that any thermodynamic irreversible processes can be described in term of generalized affinities (forces) and generalized fluxes. The product of these generalized thermodynamic forces and fluxes gives the entropy production. For example, the driving force of matter diffusion is $\nabla(\mu/T)$, when there is a gradient of the chemical potential. As a consequence, a flux of matter appears inside the system (Fick's law). In the same way, the driving force of heat diffusion is $\nabla(1/T)$. As a consequence a heat flow appears in the system (Fourier's law). The driving force of chemical reactions is $A/T$ rather than $A$. As a consequence, a reaction appears. The system always tends to bring back the system towards equilibrium when there is a displacement from equilibrium. This is the consequence of the principle of Le Chatelier-Braun. Generally, $\xi$ can be regarded as the advancement of an internal parameter (internal degree of freedom) of the system [41], and it can characterize, for instance, the equilibrium or the non-equilibrium of the matter repartition in the system. From De Donder's developments on chemical affinity, a generalization was made by different authors who applied this thermodynamic approach to any internal degree of freedom of a sample. This was particularly used in the study of glass transition, when relaxation time of processes becomes slow as compared to the time scale of the measurement.

*4.3. Configurational heat capacity*

To the best of our knowledge, the notion of non-equilibrium thermodynamic state due to the freezing of internal degree of freedom (the chemical equilibrium being not reached) during the glass transition was first due to Simon [67, 68]. For a good representation of the configurational heat capacity, let us paraphrase Bernal in the reference [69] page 35: "*The idea of a configurational specific heat for liquids, i.e., of absorption of energy not in activating further degrees of freedom but in changing potential energy, is necessary to explain the observed greater specific heat of all simple (and most other) liquids compared with that of the crystals and the occurrence in certain cases, e.g. water, of specific heats greater than 6k, which cannot be explained by any hypotheses depending on degrees of freedom only.*" Also, close to the De Donder's approach, let us cite Davies and Jones presenting the ideas of Simon in the reference [70] page 375: "*Simon pointed out that as a glass is cooled through its transformation temperature the molecular diffusion which is necessary to effect the appropriate change in configuration is increasingly inhibited and finally becomes practically impossible. Thus the value of z will become fixed somewhere near the transformation temperature and that part of the specific heat corresponding to changes in potential energy will be eliminated below this temperature. The 'configurational' contribution to any other property will similarly disappear. At the same time the system ceases to be in true internal thermodynamic equilibrium.*" In this reference, $z$ is equivalent to the order parameter $\xi$ in the present paper. This notion of configurational heat capacity was very well explained by Kauzmann in the reference [71] (section B called "*Equilibrium and dynamic mechanisms in the glass transformation*"). For a general approach of non-equilibrium thermodynamic coefficients and particularly heat capacity by means of the affinity of De Donder, see also the book of Frenkel [72], and Prigogine and Defay [59, 62]. Now, we envision the notion of non-equilibrium heat capacity in details with Prigogine and Defay.

*4.4. Configurational heat capacity of Prigogine and Defay*

*4.4.1. Non-equilibrium heat capacity*

Prigogine and Defay pushed further the reasoning of De Donder. It is indeed possible to derive a general formula of the heat capacity during a non-equilibrium calorimetric experiment. Indeed, for an irreversible calorimetric experiment, $\xi$ has no time to reach its equilibrium value $\xi_{eq}$, because of the non-zero value of the kinetic relaxation time constant $\tau$, of the transformation. In this case, the variation of the affinity is different from zero (equation (24)) and:

$$\frac{d\xi}{dT} = \frac{\alpha}{\beta} - \frac{dA}{\beta dT} \quad (34)$$

Replacing this expression in the fundamental equation (28) yields:

$$C_{mes} = C_{\xi} + \partial H / \partial \xi)_T \left[ \frac{\alpha}{\beta} - \frac{dA}{\beta dT} \right] \quad (35)$$

which can be more explicitly written as the following:

$$C_{mes} = C_{\xi} + \frac{[\partial H / \partial \xi)_T]^2}{T\partial^2 G / \partial \xi^2)_T} + \frac{A\partial H / \partial \xi)_T}{T\partial^2 G / \partial \xi^2)_T} - \frac{\partial H / \partial \xi)_T}{\partial^2 G / \partial \xi^2)_T}\frac{dA}{dT} \quad (36)$$

The formula (36), derived in 1946 by Prigogine and Defay (equation (26.84) page 121 of the reference [62]) is the general formula of the measured apparent heat capacity during an irreversible calorimetric experiment. It is a fundamental equation in the field of dynamic calorimetry. It may be applied near and far from equilibrium. The three last terms of the right hand side of the equation (36) constitute the configurational heat capacity of Prigogine and Defay. These three terms take into account the equilibrium or non-equilibrium behavior of the degree of advance of any internal degree of freedom inside the sample.

If the heat rate supplied to the sample is very large, then it is possible that the degree of advance of the transformation does not change during the time interval $\Delta t$. It is, for a given value of the step $\Delta T$, the largest irreversible experiment. The internal degree of freedom represented by the degree of advance $\xi$, is completely frozen. In this case, the total differential of the affinity (24) becomes:

$$\frac{dA}{dT} = \alpha \quad (37)$$

and with (36) the measured heat capacity becomes:

$$C_{mes} = C_{\xi} \quad (38)$$

Hence, for the largest irreversible experiment, the measured apparent heat capacity is equal to the true heat capacity of the system. With this approach, $C_{\xi}$ is the heat capacity composed by the infinitely fast degrees of freedom of the sample as compared to the time scale $\Delta t$ of the measurement. It is experimentally observed in glass transitions because of the large value of the kinetic relaxation times. To our point of view, it is for the same reason that sometimes this effect is observed in ac-calorimetry experiments when the frequency becomes large compared to the kinetic relaxation time of the process under study [73].

For an intermediate irreversible calorimetric experiment, $\xi$ has an intermediate value between 0 and $\xi_{eq}$. Hence, the measured apparent heat capacity has an intermediate value between $C_\xi$ (true heat capacity) and $C_{rev}$ (true heat capacity plus the total contribution of the heat of reaction at equilibrium). In ac-calorimetry measurements, this intermediary regime is frequently observed [22, 74, 75].

#### *4.4.2. Non-equilibrium thermodynamics close to equilibrium*

##### *4.4.2.1. Assumptions of the linear response in thermodynamics*

There are three different regimes in thermodynamics. The first is the regime of classical equilibrium thermodynamics, principally developed by Gibbs (see reference [44]) and largely spread in the literature. The second is the regime of non-equilibrium thermodynamics near equilibrium (linear regime). The thermodynamic variables never move far from equilibrium, and they can be linearized around their equilibrium values. Relaxations towards equilibrium are simple exponential relaxations. The third is the regime far from equilibrium governed by non-linear behaviors of the variables. This regime was well described by Glansdorff and Prigogine [66]. In this paper, we deal with the first and the second regime. It is difficult to find precise criteria defining the linear regime for irreversible calorimetric experiments. In TMDSC the subject was well treated by lots of authors [29, 76-81]. The qualitative criterion that we used here is, that during the finite time interval $\Delta t$, the temperature increment $\Delta T$ is not so high that even in the extreme irreversible case ($\Delta\xi = 0$ during $\Delta t$), the degree of advance will never be far from its equilibrium value. In fact, the determination of the linearity range around the state of equilibrium depends on the physico-chemical event under study.

In this linear regime, three important assumptions can be pointed out:

- De Donder and others have demonstrated that near equilibrium there is a simple proportional relation between the affinity and the rate of reaction [82-84]:

$v = aA$ (39)

where $a$ is a positive coefficient (the angular coefficient defined by De Donder) which depends only on the physical variables of the system. In our case $a = a(T)$. The formula (39) can be understood intuitively because close to the reversible transformation the affinity and the reaction rate tend together towards zero. More generally, in non-equilibrium thermodynamics close to equilibrium, there is always a proportional link between forces and fluxes present in the system (Onsager relations). Let remark that in certain case, the proportional relation holds even for high variations of the thermodynamic forces. For example, the Fourier's law remains valid even for large $\Delta T$. Anyway, Prigogine and co-workers showed experimentally that this assumption is exact for a certain number of chemical reactions [83].

- The second assumption is that, near equilibrium, the heat of reaction and the second derivative of the free Gibbs energy are close to their values at equilibrium:

$\partial H/\partial\xi)_T = \partial H/\partial\xi)_T^{eq}$ (40) and $\partial^2 G/\partial\xi^2)_T = \partial^2 G/\partial\xi^2)_T^{eq}$ (or $\beta = \beta_{eq}$) (41)

which is equivalent to neglect the second derivative of the enthalpy and the third derivative of the free Gibbs's energy with respect to the degree of advance of the transformation. This assumption is equivalent to average the variables under interest ($(\partial H/\partial\xi)_T$ and $(\partial^2 G/\partial\xi^2)_T$ ) on the small considered temperature interval $\Delta T$ around the equilibrium state defined by $T_{dc}$ and $\xi_{eq}$. This is much easier to fulfill if these variables do not vary a lot over the considered temperature interval.

- The third is certainly the hardest to justify. It is assumed that the affinity is negligible as compared to the heat of reaction:

$A << (\partial H/\partial\xi)_T$ (or $\alpha = \alpha_{eq}$) (42)

With the Berthelot-De Donder's formula (26), we see that, for a given heat of reaction, this inequality is easier to fulfil if the absolute temperature is increased. Inversely, at low temperature the measurement of the heat of reaction (with calorimetry for example) gives directly the affinity with a good approximation. This assumption is the same that the well-known approximation, $A << RT$, in chemistry. However, from a pure theoretical aspect, it is always possible to find an area close to equilibrium where these three assumptions are fulfilled.

*4.4.2.2. Non-equilibrium heat capacity close to equilibrium*

From the equation (28), if the heat of reaction is replaced by its equilibrium value and if the rate of reaction takes its value from (39), then the formula of the heat capacity measured during an irreversible calorimetric experiment near equilibrium is written as follows:

$$C_{mes} = C_{\xi} + \frac{\partial H / \partial \xi)_T^{eq} aA}{dT / dt} \quad (43)$$

This equation was first derived by Prigogine and Defay in the reference [59] (equation (19.19) page 307).

*4.4.2.3. Entropy production close to equilibrium*

At this level, we emphasize that in the equation (43), the term of entropy production was already neglected. Indeed, from the Berthelot-De Donder's formula the assumption, which consists to take the equilibrium value of the heat of reaction near equilibrium, is equivalent to the third assumption. Consequently, the third term of the right hand-side of the equation (33) ("uncompensated heat capacity of Clausius") was neglected as compared to the second term. In other words, the entropy production is neglected near equilibrium. This can be also demonstrated in an other way: near equilibrium, the rate of reaction is proportional to the affinity (first order in $A$). Hence, from (19) the entropy production is written as follows:

$$\sigma_i = \frac{a}{T} A^2 \quad (44)$$

which is of second order in $A$ and negligible.

*4.4.2.4 Total differential of the affinity close to equilibrium*

Taking into account the three last assumptions, the differential equation governing the affinity when the thermodynamic transformation occurs near equilibrium is obtained from the equation (24):

$dA/dt + \beta_{eq} aA = \alpha_{eq} dT/dt$ (45)

We define the relaxation time constant of the affinity by:

$\tau = 1/\beta_{eq} a$ (46)

It represents the kinetic time constant of the irreversible process occurring during a non-equilibrium thermodynamic transformation. This relaxation time is positive ($\beta_{eq} > 0$ and $a > 0$). Near equilibrium, the linear relation of Onsager gives:

$$v = L\frac{A}{T} \quad (47)$$

where $L$ is the Onsager's phenomenological coefficient. With this linear relation, the relaxation time constant of the affinity can be written:

$$\tau = \frac{T}{\partial^2 G/\partial\xi^2)_T^{eq} L}$$ (48) (See references [85] and [86])

Thus, near equilibrium the fundamental differential equation governing the affinity is of first order with a forcing term containing the temperature rate:

$$\tau dA/dt + A = \tau\alpha_{eq}dT/dt \quad (49)$$

Hence, knowing the temperature rate it is possible to derive the time dependent law of the affinity just by means of a simple first order differential equation. This fact is interesting for calorimetrists who know and control in general the temperature program and temperature rate. This differential equation was also derived by Prigogine and Defay in the reference [59]. The two authors have seen very well that, knowing the time dependant affinity, it is possible to derive the time dependant heat capacity from the equation (43). In 1998, Baur and Wunderlich used nearly the same approach for directly derived the so-called formula of the complex heat capacity during TMDSC experiments [87, 88]. Before entering in the details of this original approach, let make an interesting remark. If we integrate the total differential of the affinity (24) on the time interval $\Delta t$, then we obtain (assuming that the initial state at $t = 0$ is an equilibrium state):

$$A(\Delta t) = \alpha_{eq}\Delta T - \beta_{eq}\Delta\xi \quad (50)$$

with

$$\Delta\xi = \xi(\Delta t) - \xi_{eq}(0) \quad (51)$$

If the transformation is an equilibrium transformation then $A(\Delta t) = 0$ and:

$$\Delta T = \frac{\beta_{eq}}{\alpha_{eq}}\Delta\xi_{eq} \quad (52)$$

with

$\Delta\xi_{eq} = \xi_{eq}(\Delta t) - \xi_{eq}(0)$ (53)

Assuming that $\Delta T$ is constant in the different types of experiment (only $\Delta t$ is changed), then (50) becomes:

$A(\Delta t) = \beta_{eq}\Delta\xi_{eq} - \beta_{eq}\Delta\xi = \beta_{eq}(\xi_{eq}(\Delta t) - \xi(\Delta t))$ (54)

Close to equilibrium, the affinity is proportional to the distance of the degree of advance of the transformation from its equilibrium value. If $\xi < \xi_{eq}$, then the affinity is positive, and if $\xi > \xi_{eq}$, then the affinity is negative. If the affinity is positive then the rate of reaction is positive because of the fundamental inequality of De Donder (16). Thus, the degree of advance of the transformation increases. If the affinity is negative then the rate of reaction is negative. Thus, the degree of advance of the reaction decreases. After a perturbation, the transformation always drives the system towards equilibrium. This is the consequence of the principle of Le Chatelier-Braun. The equation (54) was directly used by Claudy and Vignon who have derived the distance of the degree of advance of the transformation from the equilibrium, $\Delta\xi = \xi - \xi_{eq}$, in the time interval $\Delta t$ in order to explain the complex heat capacity in TMDSC [89]. In their article, the coefficient $k(T)$ is equal to the inverse of the kinetic relaxation time $\tau$ used in this paper. The equation (54) can be rewritten in the following form:

$A = (\xi_{eq} - \xi)/a\tau$ (55)

and with the equation (39), the rate of reaction becomes simply:

$v = (\xi_{eq} - \xi)/\tau$ (56)

## 5. Generalized calorimetric susceptibility

In two recent publications dating from 1998, Baur and Wunderlich used the previous thermodynamic approach of De Donder-Prigogine-Defay in order to directly derive the formula of the complex heat capacity [87, 88]. In their article, they used this thermodynamic approach with the purpose of seeing whether the notion of complex heat capacity is meaningful in TMDSC measurements. On our point of view, we think that this approach is original and however can be a complementary manner in order to have access to the physical meaning of the frequency dependent complex heat capacity as compared to the usual linear response theory approach. Other scientists have already used macroscopic non-equilibrium thermodynamics in calorimetry, generally but not necessarily in the calorimetric study of glass transitions [24, 89-93]. Before enter in the details of the derivation of Baur and Wunderlich, we would like to show with the help of few references that the notion of frequency dependent complex heat capacity has been already used a long time ago, although it was not particularly connected to the field of calorimetry.

### *5.1. Ultrasonic absorption*

To our knowledge, the first time that the notion of frequency dependent complex heat capacity appeared in the literature, was at the beginning of the 20$^{th}$ century in scientific works concerning the propagation of sound in different mediums. A very detailed review on the subject has been written by Alig [94]. In the propagation of sound, the oscillation of acoustic pressure is coupled with an adiabatic temperature oscillation. Relaxation phenomena inside the material provide a dispersion and an absorption of the sound wave. This effect is explained by the existence of a complex heat capacity for which the imaginary part, linked to the absorption, reflects a problem of energy transfer between internal and external degrees of freedom. In an interesting review on "Supersonic Phenomena" written by Richards in 1939, the notion of frequency dependent heat capacity is also treated [95]. In this article, the distinction between low frequency heat capacity, $C_0$, and high frequencies heat capacity, $C_\infty$ is already made. Considering only a single transition $0 \leftrightarrow 1$ between two degrees of freedom, Richards used a reasoning taking into account a principle of microscopic reversibility (or a principle of detailed balance), considering the probabilities of transition between the two states, in order to derive the formula (1) of the frequency dependent complex heat capacity. Jeong used the same type of reasoning in his review in dynamic calorimetry to derive the formula (1) with the help of two different temperatures (one is fictive) for characterizing the internal (slow) degrees of freedom and the external (fast) degrees of freedom (phonon bath) [28]. In his review, Richards said that the first indication that dispersion due to heat capacity could be expected was due to Jeans in 1904 [96] (although we have not found that in the French edition book dating from 1925).

### *5.2. Generalized calorimetric susceptibility of Davies.*

It was in 1956 in a publication written by Davies, that for the first time the notion of complex thermodynamic quantities and more specifically complex heat capacity was derived directly from the thermodynamics of De Donder, Prigogine and Defay [97]. This interesting paper begins with a clear historical introduction of non-equilibrium thermodynamics. Lots of references on the subject can be inferred from this paper. The tight link existing between regression in time of microscopic fluctuations and non-equilibrium relaxations of macroscopic thermodynamic variables are well discussed. From this point of view, we can say that perturbing a system by modulating its temperature is equivalent to provoke macroscopic temperature fluctuations of the system. The part II of this paper entitled "*Incomplete system under uniform conditions: relaxation*" is a presentation of the De Donder's approach of chemical reactions. Then, with the help of an order parameter $z$ (equivalent to $\xi$ in our manuscript), the general approach of static and dynamic thermodynamic coefficients such as $C_p$ (heat capacity at constant pressure), $\alpha$ (thermal expansion coefficient at constant pressure) and $\kappa$ (modulus of compressibility at constant temperature) was considered. The names of Frenkel, Prigogine, Defay and Meixner were often cited. Davies also made the distinction between frozen coefficients (such as $\partial V/\partial T)_{p,z}$ ) in which the reaction is not allowed to proceed with a fixed value of the order parameter, and equilibrium coefficients (such as $\partial V/\partial T)_{p,A}$ ) which are evaluated at constant affinity. Next, Davies used a very general formalism of this approach with undefined order parameters and matrix treatments. In this part, Davies defined dynamic thermodynamic coefficient as follow: "*It is usually specified by means of an impedance function connecting the Fourier transforms of X and x*" ($X$ and $x$ being thermodynamic conjugate variables). When $Q$ (or better $S$) and $T$ are taken as thermodynamic conjugate variables, this dynamic coefficient is the so-called complex heat capacity. He also envisaged the case of a continuous distribution of order variables and relaxation times. After, Davies applied this approach to the treatment of glass transition where he assumed that a single ordering parameter is sufficient to define the transition.

### *5.3. Generalized calorimetric susceptibility of Eigen and De Mayer*

In the volume VIII, part II, of the Technique of Organic Chemistry, called "*Investigation of Rates and Mechanisms of Reactions*", Eigen and De Mayer wrote a very detailed paper on theoretical and experimental techniques about chemical relaxation [98]. This paper could be of interest for calorimetrists because the two authors envisaged with a lot of details the methods consisting in perturbing the chemical equilibrium of a system by means of temperature variations. Although calorimetry was not explicitly mentioned in this paper, they reviewed the so-called "temperature jump method" very used in chemistry and biology, which consists to perturb rapidly the temperature of a system and to record a physical parameter (such as optical absorption) without necessary measuring the heat resulting of this perturbation. After giving the theoretical basis of relaxation methods, Eigen and De Mayer investigated relations between thermodynamics and relaxation times. Again the names of De Donder, Meixner or Davies were extensively cited. Then, an interesting development of relaxation based on stationary methods, but treated on a thermodynamic point of view, was given. In the section called "*Entropy-Temperature*" they fund again the De Donder's formula of heat capacity measured at constant affinity (see equation (29)). Next, in a section called "*dynamic equations of state*" taking into account a stationary perturbation of the system around its equilibrium state, they derived explicitly all the dynamic complex frequency dependent thermodynamic variables, such as the coefficients of isothermal and adiabatic compressibility and the complex specific heat at constant volume or pressure. These derivations were particularly applied to chemical equilibrium, but as we have already mentioned, they were more largely applied to any order parameter concerning peculiar degrees of freedom within a sample having their own thermodynamic relaxation times.

*5.4. Generalized calorimetric susceptibility of Baur and Wunderlich*

In 1998, using macroscopic non-equilibrium thermodynamics of De Donder, Prigogine and Defay, Baur and Wunderlich derived for the first time in calorimetry (TMDSC) the well-known formula of the complex heat capacity [87, 88]. This treatment is also described in the recent book of Wunderlich [99]. Before enter in the detail of this derivation, it is important to give two precisions. Firstly, in the literature dealing with thermodynamics applied to calorimetry, it is rather usual to encounter an erroneous basic definition of the heat capacity. The mistake was also made in the article of Eigen and De Mayer. The measured heat capacity is generally wrongly defined as follows:

$C_p = TdS/dT)_p$ (57)

Indeed, the exact definition is slightly different and given by the equation (28):

$C_p = \delta Q/dT = dH/dT)_p$ (28)

The equation (57) is equivalent to (28) only at thermodynamic equilibrium or near thermodynamic equilibrium ($A$ is neglected compared to the value of the heat of reaction). This mistake has no consequence for the derivation of the complex heat capacity measured near equilibrium, because the entropy production (negligible near equilibrium) is forgotten in this definition (see discussion in the section *4.4.2.3.* called "*Entropy production close to equilibrium*"). Indeed, the measured heat capacity is only linked to the entropy exchanged (heat exchanged) between the system and the surroundings:

$C_p = Td_eS/dT = TdS/dT - Td_iS/dT$ (58)

Secondly, the conclusion given by Baur and Wunderlich in their paper seems to us very pessimistic. They concluded that the notion of complex heat capacity is not very useful in TMDSC measurements. Perhaps complex heat capacity is indeed not adapted for TMDSC experiments because of parasitic effects, which has to be taken into account, such as non-adiabaticity and thermal contact between samples and sensors. These unwanted effects can indeed induce other relaxation times and parasitic frequency and imaginary components. On the other hand, this notion can be very useful in ac-calorimetry measurements where thermal equilibrium conditions (adiabaticity and homogeneity of the temperature) are generally respected (with the use of adiabatic plateau) and in other dynamic methods as fast speed DSC if the two last conditions are fulfilled [100-102]. To our point of view, we think at contrary that the derivation made by Baur and Wunderlich of the complex heat capacity is unusual and original as compared to the linear response theory. Also, this approach can provide a new regard in dynamic calorimetry field and can give a better physical understanding of frequency dependent complex heat capacity. To our point of view, the only restriction concerning usefulness of the complex heat capacity is the respect of linearity and stationarity criteria. In ac-calorimetry these two conditions are easier to fulfill if the amplitude of the temperature oscillation is small and the rate of the mean temperature is low. In the next, we assume that thermal equilibrium conditions, linearity and stationarity conditions are respected.

Let take the case of ac-calorimetry experiments. In the range of working frequency defined by the two following inequalities:

$\tau_{int} << 1/\omega << \tau_{ext}$ (59)

which define the strict conditions of thermal equilibrium for the heat capacity measurement, the temperature oscillation of the sample is written:

$T_{ac} = \delta T_{ac} \cos(\omega t - \pi/2)$ (60)

with $\delta T_{ac} = P_0/\omega C_{mes}$ (61) and $\varphi = \pi/2$ (62)

$\varphi$ is the phase lag of the modulated temperature as compared to the input oscillating thermal ac-power $P_0 \cos(\omega t)$. $\tau_{int}$ is the internal relaxation time of the temperature which takes into account the thermal diffusivity within the sample and the thermal interface conductance between the thermometer, the heater and the sample. $\tau_{ext}$ is the external relaxation time of the temperature towards the thermal bath of temperature $T_0$. Let insist that temperature oscillations occur around a mean temperature assumed to be constant $T_{dc}$, which defines the equilibrium sate $(T_{dc}, \xi_{eq})$. It is the stationary condition of the measurement. It occurs when $T_{dc}$ is maintained constant (measurement step by step waiting for the equilibrium) or when it varies very slowly (slow ramp). Now, what happens when a physico-chemical transformation with a finite kinetic time constant arises? By means of the linearity assumption, we assume that in the presence of a physico-chemical transformation in the sample, the temperature oscillates with the same frequency that when there is no transformation. Hence, the transformation modifies only the amplitude and the phase of the oscillating temperature, which can be written:

$T_{ac} = \delta T_{ac} \cos(\omega t - \pi/2 - \kappa)$ (63)

where $\kappa$ is the phase lag generated by this non-equilibrium effect. The temperature rate is:

$dT/dt = dT_{ac}/dt = i\omega T_{ac}$ (64)

The dc temperature or the mean temperature, $T_{dc}$, is obviously not involved in the time derivative, $dT/dt$.

At this level, the original idea of Baur and Wunderlich was to exactly derive the differential equation driving $\xi$ from the total differential of $A$ (24) using the linear relation (39) between $v = \dot{\xi}$ and $A$ (the dot on the variable represents the time derivative). This yields a non-linear second order differential equation in the variable $\xi$, where the forcing term contains the temperature rate (equation (19) of the reference [88]). After a rather complex calculus, they linearized the solution of this differential equation and with (64), they found the expression of $\dot{\xi}$ in function of the affinity and the other parameters of the equation (equation (23) of the reference [88]).

More simply, starting with the help of the assumptions of the linear regime, the equation (49) driving $A$ is simply written in the oscillatory regime as the following:

$\tau dA/dt + A = i\omega\tau\alpha_{eq}T_{ac}$ (65)

It is simply resolved taking into account all the assumptions made close to equilibrium (all the temperature dependant variables are assumed to be constant around the equilibrium state over the amplitude $\delta T_{ac}$):

$$A = \frac{i\omega\tau\alpha_{eq}T_{ac}}{1 + i\omega\tau} \quad (66)$$

which is equivalent to the sus-cited equation (23) of the article of Baur and Wunderlich. Consequently, in the case of modulated temperature experiments, the affinity, which is the response of an oscillating temperature, is also an oscillating function with two components, one being in-phase and the other being out-of-phase. The amplitude of each component depends on the value of the ratio, $\omega\tau$, as compared to the unity.

Let more explicitly rewrite the affinity as follow:

$$A = \partial H / \partial \xi)_T^{eq} \frac{T_{ac}}{T_{dc}} \frac{(\omega\tau)^2}{[1+(\omega\tau)^2]} + i\partial H / \partial \xi)_T^{eq} \frac{T_{ac}}{T_{dc}} \frac{\omega\tau}{[1+(\omega\tau)^2]} \quad (67)$$

When the irreversibility is very low, that is to say when $\omega\tau << 1$, then:

$$A = i\partial H / \partial \xi)_T^{eq} \omega\tau \frac{T_{ac}}{T_{dc}} \quad (68)$$

which tends to zero at the limit of the reversible experiment ($\omega\tau = 0$). When the irreversibility is maximum (arrested equilibrium), this is to say $\omega\tau >> 1$ (at the limit $\omega\tau \to +\infty$) then:

$$A = \partial H / \partial \xi)_T^{eq} \frac{T_{ac}}{T_{dc}} \quad (69)$$

The affinity can also be written in another way:

$$A = \delta A exp(i\phi) T_{ac} \quad (70)$$

where:

$$\delta A = \frac{\omega\tau\alpha_{eq}}{\sqrt{1+(\omega\tau)^2}} = \frac{\alpha_{eq}}{\sqrt{1+1/(\omega\tau)^2}}$$ (71) and $\phi = arctg(1/\omega\tau)$ (72)

If $\omega\tau = 0$ then $\phi = \pi/2$ and if $\omega\tau = +\infty$ then $\phi = 0$. Finally:

$$A = \partial H / \partial\xi)_T^{eq} \frac{T_{ac}}{T_{dc}\sqrt{1+1/(\omega\tau)^2}} \exp(i\phi) \quad (73)$$

The phase difference between the affinity and the oscillating temperature is $\phi$, which is equal to $\pi/2$ for the reversible experiment and $0$ for the maximal irreversible experiment. The oscillating affinity and temperature are represented with their respective components in the Fresnel's diagram of the figure 3.

Now, as Baur and Wunderlich did, by means of the equation (70) for example, the formula of the frequency dependent complex heat capacity can be easily derived from the equation (43) of the non-equilibrium heat capacity near equilibrium:

$$C_{mes} = C_\xi + \frac{\eta_{eq}\exp(-i\psi)}{\sqrt{1+(\omega\tau)^2}} \quad (74)$$

with $\psi = \frac{\pi}{2} - \phi = arctg(\omega\tau)$ (75)

This formula can be more explicitly rewritten:

$$C_{mes} = C_{\xi} + \frac{\eta_{eq}}{1 + i\omega\tau} \quad (76)$$

Knowing that $C_{\xi} = C_{mes}(\omega = +\infty)$ and $C_{rev} = C_{mes}(\omega = 0)$, (76) is exactly the formula (1) of the complex heat capacity.

Therefore, by means of the affinity, from the formalism of De Donder-Prigogine-Defay on irreversible thermodynamics, the well-known formula of the complex heat capacity was directly derived from the two principles of thermodynamics. The frequency dependent complex heat capacity is thus the consequence of irreversible thermodynamics near equilibrium in the linear regime. *C'* and *C''* are due to the generation of an oscillating affinity with in-phase and out-of phase components during non-equilibrium physico-chemical transformations. This is due to the non-zero value of the ratio, $\omega\tau$.

Let now point out one important remark. As we have already mentioned in the foregoing, the entropy production (to be precise the thermal power of irreversibility $P_i$) was neglected near equilibrium. Consequently, the generalized calorimetric susceptibility is not directly the consequence of the thermal power due to the internal entropy production within the sample when it is perturbed near equilibrium. In other words, surprisingly, the uncompensated heat of Clausius does not disturb the measured heat capacity during dynamic calorimetric experiments, which was not obvious beforehand. It is certainly not the case when the calorimetric experiment goes outside the linear regime far from equilibrium.

Then, Baur and Wunderlich discuss the influence of the two following extreme cases on *C'* and *C''*. Firstly, at the limit of the reversible experiment (internal equilibrium, $\omega\tau \to 0$):

$C' = C_{rev}$ and $C'' = 0$ (77)

At the limit of the maximum irreversible experiment (arrested equilibrium, $\omega\tau \rightarrow +\infty$ ):

$C' = C_\xi$ and $C'' = 0$ (78)

When $\omega\tau$ is of the order of the unity, $C''$ is a maximum and $C'$ is contained between the two previous extreme cases (intermediary regime). Hence, thermodynamic irreversibility of a peculiar degree of freedom inside the sample is the explanation of the frequency dependent heat capacity effect measured in ac-calorimetry experiments close to equilibrium [22, 73-75].

Finally, Baur and Wunderlich derived for the first time, in the general case, the so-called formula of the entropy produced by an irreversible process during non-equilibrium calorimetric measurements. We remember that the affinity oscillates with a phase advance of $\phi$ compared to the oscillating temperature. Hence, the affinity is a real number, which is explicitly written without complex notations:

$A = \delta A \delta T_{ac} cos(\omega t - \pi/2 - \kappa + \phi)$ (79)

With (20), (44) and (71) the power of irreversibility is written as the following:

$$P_i = \eta_{eq} \frac{\delta T_{ac}^2}{T_{dc}\tau[1+1/(\omega\tau)^2]} \cos^2(\omega t - \pi/2 - \kappa + \phi) \quad (80)$$

Integrating this expression over one period of the oscillating temperature, the time-averaged irreversibility power is:

$$\overline{P}_i = \pi\eta_{eq}\frac{\delta T_{ac}^2\omega\tau}{T_{dc}[1+(\omega\tau)^2]} = \pi\frac{\delta T_{ac}^2}{T_{dc}}C'' \quad (81)$$

The instantaneous irreversible entropy production is:

$$\sigma_i = \eta_{eq}\frac{\delta T_{ac}^2}{T_{dc}^2\tau[1+1/(\omega\tau)^2]}\cos^2(\omega t - \pi/2 - \kappa + \phi) \quad (82)$$

Over one period of the oscillating temperature, the time-averaged irreversible entropy production (or simply mean entropy production) is given by:

$$\overline{\sigma}_i = \pi\frac{\delta T_{ac}^2}{T_{dc}^2}C'' \quad (83)$$

which is the formula found by Baur and Wunderlich, but in their paper the mean entropy production is taken over half-period of the oscillation. This expression was approximately already derived in the literature, but only in a peculiar case, as we shall see in a next section. Knowing that the modulus of the oscillating temperature is written:

$$\delta T_{ac} = \frac{P_0}{\omega|C_{mes}|} \quad (84)$$

(83) is written as follows:

$$\overline{\sigma}_i = \pi\frac{P_0^2}{\omega^2 T_{dc}^2}\frac{C''}{|C_{mes}|^2} \quad (85)$$

The amount of energy involved per half-period of the oscillation is:

$$\delta Q_0 = \int_{-T/4}^{T/4} P_0 \cos(\omega t) dt = 2\frac{P_0}{\omega} \quad (86) \; (P_0/\omega \text{ per quart-period})$$

Thus, the mean entropy production, per half-period of the oscillation, of the irreversible processes occurring within the sample during non-equilibrium calorimetric experiments is:

$$\overline{\sigma}_i = \frac{\pi}{4}\frac{\delta Q_0^2}{T_{dc}^2}\frac{C''}{\left|C_{mes}\right|^2} = \frac{\pi}{4}\frac{\delta Q_0^2}{T_{dc}^2}\mathrm{Im}\left(\frac{1}{C_{mes}}\right) \quad (87)$$

In other words, this time averaged entropy production is directly proportional to the imaginary part of the complex impedance of the measurement (see also reference [30]). For instance, we can conclude that if there is a dissipation of heat, or heat loss, in dynamic oscillating calorimetric experiments, it should be more physically linked to the imaginary part of the dynamic calorimetric impedance of the measurement and not simply to the imaginary part of the complex heat capacity. However, both the imaginary part of the complex heat capacity and the complex impedance vanish at equilibrium and at completely frozen equilibrium. Our point of view of the question will be given in the next section.

## 6. Generality of the non-equilibrium thermodynamics approach in dynamic calorimetry

In this section we would like to show that the previous approach can be regarded on a very general manner in dynamic calorimetry. Of course, the previous model is simplified. Firstly, it can be used only near thermodynamic equilibrium. Certainly, even with low temperature rates, lots of transitions should occur in the non-linear regime far from equilibrium. In this case, the affinity and the entropy production cannot be neglected in the model. Also, what happens in real first-order transition when heat capacity curves become very sharp? Secondly, to consider the freezing of one peculiar degree of freedom with one relaxation time constant can just be applied to simple chemical reaction. Nevertheless, in this case the model can be complicated considering, as Davies did, a distribution of relaxation time constant or multiple time constants [97]. In this section, firstly we will show that under the conditions previously mentioned, this approach can be usefully applied to all dynamic calorimetric experiments. The case of simplified classical DSC is treated. Some comments will be given for the study of the glass transition via this model. Secondly, we would like to show that the derivation of the so-called formula of the entropy production (equation (83)), generally derived in the literature, is only a peculiar case of the derivation made by Baur and Wunderlich in the general case of irreversible processes. Indeed, this is simply the irreversible process due to thermal relaxation towards the heat bath. It can be also obtained from the irreversible process due to diffusion of heat inside the sample as we will see. In the last part, we will give our point of view on imaginary part of complex heat capacity and heat dissipation in heat capacity measurements during non-equilibrium physico-chemical transformations.

*6.1. Macroscopic non-equilibrium thermodynamics applied to dynamic DSC.*

Let be a DSC experiment with a constant temperature rate:

$dT/dt = constant = \gamma$ (88)

As already mentioned, the case of a pure isothermal first-order phase transition ($\gamma = 0$) is not envisaged here. We consider at first that the experiment is realized close to equilibrium, which means that the temperature ramp is not too fast, but rather fast to unbalance the system. Let assume that in the time interval $\Delta t$, the temperature step is $\Delta T$. Then we consider that this temperature variation occurs around a constant mean temperature $T_{dc}$. Thus, for $0 < t < \Delta t$, we have $T_{dc} - \Delta T/2 < T < T_{dc} + \Delta T/2$. We also assume that $\tau$ and $\alpha_{eq}$ are constant around $T_{dc}$ and also that they have very small and smooth variations during the entire calorimetric experiment.

For $0 < t < \Delta t$, the general solution of the differential equation governing the affinity (49) is:

$$A = A_0 exp(-t/\tau) + \tau\alpha_{eq}\gamma[1 - exp(-t/\tau)] \quad (89)$$

where $A_0$ is the initial value of the affinity at the time $t = 0$.

We shall now envisage three different situations:

i) $\tau/\Delta t << 1$

In this case, for a large majority of times included into the interval $0 < t < \Delta t$, the relaxation time $\tau$ is negligible. Thus:

$$A = \tau\alpha_{eq}\gamma \quad (90)$$

On this time interval, the measured heat capacity is given by the formula (43):

$C_{mes} = C_{rev}$ (91)

This is the reversible case.

ii) $\tau/\Delta t >> 1$

For every time in the interval $0 < t < \Delta t$, the affinity is with (49):

$A = A_0$ (92)

Hence, with (43):

$$C_{mes} = C_{\xi} + \eta_{eq} \frac{A_0}{\tau \alpha_{eq} \gamma} \quad (93)$$

It is the maximum possible irreversible experiment. We see that $C_{mes} = C_{\xi}$ only if the initial state is a state of equilibrium ($A_0 = 0$).

iii) $\tau/\Delta t \approx 1$

In this case, $\tau$ and $\Delta t$ are of the same order. Including the general expression of the affinity (89) in the formula (43), it gives on the time interval $0 < t < \Delta t$:

$$C_{mes} = C_{\xi} + \eta_{eq} \left[ 1 + \left( \frac{A_0}{\tau \alpha_{eq} \gamma} - 1 \right) \exp(-t/\tau) \right] = C_{rev} - \eta_{eq} \left[ \left( 1 - \frac{A_0}{\tau \alpha_{eq} \gamma} \right) \exp(-t/\tau) \right] \quad (94)$$

This is the general formula of the measured heat capacity in a DSC experiment realized near equilibrium during non-equilibrium thermal events. The two extreme cases i) and ii) can be obtained from this equation.

For the following time intervals, the temperature ramp has brought the system to another mean temperature $T_{dc}$. Now, $\tau$ and $\alpha_{eq}$ have new values ($\tau(T_{dc})$ and $\alpha_{eq}(T_{dc})$), but with the assumption made before, we consider that they have not varied a lot. We can thus envisaged the resolution of the differential equation of the affinity in a same time interval $0 < t < \Delta t$ but around a new temperature interval $T_{dc} - \Delta T/2 < T < T_{dc} + \Delta T/2$. The three previous cases are also envisaged but with a new initial value of the affinity. This new initial affinity value depends on the following variables:

$$A_0' = A_0'(A_0, \tau(T_{dc} - \Delta T), \alpha_{eq}(T_{dc} - \Delta T), \gamma) \quad (95)$$

We may remark that in the extreme case i), the affinity depends only on the temperature dependent values of $\tau$ and $\alpha_{eq}$. It is anyway always equal to $\tau\alpha_{eq}\gamma$. Therefore, in the transition area (or thermal event area) during the DSC experiment, the heat capacity has always its equilibrium value (reversible case). On the other hand, if we consider an experiment taken in the intermediate case iii), or in the maximum irreversible case ii), the affinity depends on the value of $A_0$ which changes at each new time interval $\Delta t$. As $A_0$ changes in time (it follows simply the equation (89)), it is thus possible that before the end of the experiment it will reach the value $\tau\alpha_{eq}\gamma$ (if $\tau\alpha_{eq}$ has not varied a lot). If it is the case, we find again the reversible case and $C_{mes} = C_{rev}$. This situation can happen when the time interval of the transformation area is larger than $\tau$:

$\Delta t_{trans} = \Delta T_{trans}/\gamma >> \tau$ (96)

where $\Delta t_{trans}$ is the time duration of the thermal event area, and $\Delta T_{trans}$ the temperature interval of this transformation area.

In DSC experiment it seems paradoxical that the reversible case may be reached when the affinity becomes constant. Indeed, the system is nevertheless in a non-equilibrium state. In fact, in this case a stationary non-equilibrium state is reached and the affinity is constant along the time. Indeed, when the affinity becomes constant, there is no new affinity variation ($dA = 0$) and with (24), (34) or (35), we find again the reversible case.

Let now consider a DSC experiment with a decreasing temperature ramp, not too fast to preserve internal thermal equilibrium but fast enough to unbalance the system. If the temperature of a first order transition (liquid/solid) is crossed, bringing the system in a non-equilibrium state, the measured heat capacity will follow the equation (94). The system tries to reach its thermodynamic equilibrium state, the liquid is transformed in solid, and the time taken by the system to do that is few $\tau$. After a certain time interval, the system can transform all the liquid into solid. But let imagine that before the system may reach its state of equilibrium, its temperature attains such a value for which the relaxation time $\tau$ takes a very high value. The affinity cannot reach the value $\tau\alpha_{eq}\gamma$. The system can never reach its thermodynamic equilibrium state. It is frozen in a vitreous state defined by the equation (93). In this case, the high variation of $\tau$ with temperature causes the freezing-in of the system. The ratio $\tau/\Delta t$ becomes so high that the system is arrested in a meta-stable state. This last discussion is not very new. Since a long time ago the glass transition has been seen more as a "frozen first-order transition" than a new type of thermodynamic transition. However, these last developments in dynamic DSC by means of the macroscopic non-equilibrium thermodynamic approach of De Donder-Prigogine-Defay can be an interesting starting point. For more information on a non-equilibrium thermodynamic approach of the glass transition by means of the heat capacity measurements, see the interesting following references [103-107]. The irreversible thermodynamics approach is also used for the study of the glass transition by means of the hole theory [108-110].

*6.2. Averaged entropy production over one period of the temperature cycle.*

In the references [28, 30, 92] sus-cited, the equation (83) (or a close expression) of the entropy production averaged over one period of the temperature oscillation is always derived from the following integral:

$$\overline{\sigma}_i = \int_{-T/2}^{T/2} \frac{\dot{Q}(t)}{T(t)} dt \quad (97)$$

where $T$ in the integral limits is the period of the modulated temperature. Indeed, from this equation, taking $T(t) = T_{dc} + T_{ac}$, knowing that $\dot{Q}(t) = P_0 \cos(\omega t)$ and keeping only the term of the second order in the entropy, the equation (83) can be easily derived (see reference [30]). In these three different publications, the authors explain the existence of the imaginary part of the complex heat capacity as the consequence of the entropy exchanged by the system with the heat bath for one period of the oscillation. Höhne has emitted strong doubts about the validity of this reasoning [33]. He proposed to use rather irreversible thermodynamics to resolve the problem. In fact the formula (83) is valid, but we agree with Höhne that in this case the derivation of this formula have nothing to do with generalized calorimetric susceptibility given by the equation (1). In fact, in this case, where TMDSC method is treated, the entropy production is simply due to the non-equilibrium behavior of the oscillating temperature of the sample as compared to the temperature of the heat bath and indeed, it may have an entropy exchange over one period of the cycle between the sample and the bath. This entropy results on the non-adiabatic behavior of the TMDSC method. In TMDSC, the measurement being non-adiabatic, the system cannot be considered as isolated, and the heat bath has to be taken into account in the balance of the entropy produced. In other words, the calculated averaged entropy production is only due to the irreversible effect of the external non-equilibrium behavior of the temperature of the system as compared to the bath. In ac-calorimetry, the situation is different because the condition of adiabaticity is generally respected. Thus, the only possible entropy exchanged between the system and the heat bath is the constant heat flow, which maintains constant the mean temperature of the sample (first order term of the entropy in (97)). There is no entropy exchange due to the oscillating temperature between the sample and the heat bath. The sample alone has to be taking into account in the balance of the entropy produced. But, let imagine that the second condition of validity of ac-calorimetry measurements (internal thermal equilibrium) is not fulfilled. In this case, a heat diffusion effect occurs due to the oscillating temperature of the sample. The generalized De Donder's approach is then applied as follows. The thermodynamic driving force (generalized affinity) in presence of a temperature gradient is:

$A = \Delta(1/T)$ (98)

The generated flux is simply the heat flow propagating through the sample:

$v = \dot{Q}(t) = dQ/dt$ (99)

For a temperature oscillation of amplitude $\delta T_{ac}$ around $T_{dc}$, the driving force is written:

$$\Delta(1/T) = 1/T_{dc} - 1/(T_{dc} + T_{ac}) \approx T_{ac}/T^2_{dc} \quad (100)$$

where the amplitude of the temperature modulation $T_{ac}$ is neglected as compared to $T_{dc}$ (linear regime). Hence, the averaged entropy production over one period of the temperature modulation due to the irreversible heat diffusion effect, which is directly given by the integral of the product of the force with the induced flux, is given by:

$$\bar{\sigma}_i = \int_{-T/2}^{T/2} \dot{Q}(t) \frac{T_{ac}}{T_{dc}^2} dt \quad (101)$$

which is equivalent to the second order term of (97) with the same consequence as above. Nevertheless, the two last examples of irreversible thermal effects are peculiar cases of irreversible processes and have nothing to do with frequency dependent complex heat capacity. As Höhne well saw and Baur and Wunderlich did, the exact way to obtain (83), in the case of generalized calorimetric susceptibility, is the developments made in the reference [88] or in a closed way in the present paper.

*6.3. Physical meaning of the imaginary part of the generalized calorimetric susceptibility*

The time-averaged entropy production over one period of the temperature oscillation in modulated non-equilibrium ac-calorimetry experiments is directly proportional to the imaginary part of the generalized calorimetric impedance of the measurement. Only this imaginary part and not the imaginary part of the frequency dependent complex heat capacity could have the physical meaning of heat dissipation. Heat dissipation is generally linked to a lost of work. In dynamic calorimetry, under the assumptions made before, since no work is involved in the experiment, where is passed the "bad heat", knowing that the system returns exactly in the same state and that heat cannot have time to relax towards the heat bath (adiabaticity condition) after one period of the temperature oscillation. Once again, we try to give a beginning of explanation referring us to the work of Prigogine. In an article published in 1953, Prigogine and Mazur envisaged a general extension of thermodynamics of irreversible processes applied to systems with internal degrees of freedom [111]. They defined an internal space of configuration of the system, where each degree of freedom is represented by a continuing variable representing a coordinate of the internal space. Applying the non-equilibrium thermodynamics of continuous systems, they envisaged diffusion along the internal coordinate, which participates to the entropy production. Hence, by analogy with heat diffusion where heat is lost along the spatial direction defined by the hot and the cold points, in this representation, heat is lost by diffusion along the coordinate defined by the degree of freedom inside the internal space of configurations. Hence, for example in this model, chemical reactions can be regarded as diffusion along an internal coordinate (degree of advance of the reaction) between two stable constituents separated by a potential gap. During this diffusion effect, heat is lost (dissipated or absorbed), entropy is produced, and the mean entropy production per cycle of the oscillation is proportional to the imaginary part of the inverse of the complex heat capacity.

## 7. Conclusion

In this paper, dynamic calorimetric experiments have been envisaged by means of thermodynamics of irreversible processes initiated by De Donder in the twenties of the $20^{th}$ century. On an historical basis, we have shown that macroscopic non-equilibrium thermodynamics can be helpful for the interpretation of dynamic calorimetric measurements. After having provided definitions of dynamic calorimetry and macroscopic non-equilibrium thermodynamics, the notion of irreversible calorimetric experiments has been envisaged on a qualitative way. A focus has been made on the fact that irreversibility in calorimetric experiments is not an absolute notion, but it depends on the time scale of the measurements. More precisely, the ratio of the characteristic relaxation time of the event under study with the time scale of the measurement defines the strength of the thermodynamic irreversibility. The link between kinetic and non-equilibrium thermodynamics becomes clear. The time scale of the measurement is the time interval over which the perturbing parameter brings the system in a state of non-equilibrium around the stationary equilibrium state.

Next, quantitative macroscopic non-equilibrium thermodynamics of De Donder and the members of his school have been envisaged on a calorimetric point of view. Derivations of the well-known and less-known formulas of heat capacities measured during equilibrium and non-equilibrium physico-chemical transformations have been given. Some assumptions such as thermal equilibrium, constancy of the pressure, mechanical equilibrium, which can be with attention verified in lots of calorimetric experiments, have been considered. Assumptions of stationarity and linearity have been implicitly made when the thermal power provided to (or released from) the sample by the experimentalist is low enough to provide a sufficiently low temperature rate. In this case, the rate of the thermodynamic transformation is proportional to the generalized affinity. Near equilibrium, in the linear regime, the affinity can be considered to be negligible in relation to the heat of the transformation. Therefore, it follows a simple first order differential equation where the forcing term contains the temperature rate. This state function characterizes the force of an irreversible process, which tends to bring back the system to a state of equilibrium. Because of the intrinsic physical kinetics of the sample, this operation takes a certain amount of time. This kinetic time constant is just the relaxation time constant of the affinity. During this step, thermal power is produced within the sample, which is proportional to the affinity. The non-equilibrium measured heat capacity depends then directly on the ratio of the affinity with the temperature rate. Knowing the time dependant affinity, the time dependant heat capacity is found. At this level, the thermal power due to the positive entropy produced by irreversible processes occurring inside the sample has been neglected. Indeed, this thermal power is of the second order in the affinity. As a consequence, the apparent heat capacity measured by the experimentalist, which is the ratio of the heat exchanged between the sample and the surroundings to the temperature rate, is the sum of the true heat capacity of the sample plus a term related to the heat of the transformation whose value depends on the force of the irreversibility. This added term decreases when the force of the irreversibility increases. This is the well-known decrease effect in apparent heat capacity measured during dynamic calorimetric experiments. With this last approach, perturbing now the system with an harmonic temperature oscillation, from the work of different authors during the last century, particularly from the recent work of Baur and Wunderlich, the generalized calorimetric susceptibility or frequency dependent complex heat capacity has been directly derived. Hence, for the first time a direct connection has been established between the two fundamental principles of thermodynamic and the frequency dependent complex heat capacity. This approach, slightly different than the linear response theory approach (obviously equivalent in its foundations), can give a better physical meaning of the frequency dependent complex heat capacity and its imaginary part. To our point of view, we can now affirm the following assertions:

- Frequency dependent complex heat capacity is the consequence of irreversible physico-chemical transformations occurring in the linear regime when the temperature of a sample follows a harmonic oscillation. During this irreversible process, a thermal power proportional to the affinity is produced within the sample. It is the cause of the complex heat capacity and thus, the cause of $C'$ and $C''$. The thermal power due to the entropy produced during this irreversible process ("uncompensated heat capacity of Clausius") is negligible near equilibrium and does not perturb the heat capacity measurement. In other words, frequency dependent complex heat capacity is due to the slow kinetic of an order parameter characterizing a peculiar internal degree of freedom of the sample when the temperature is harmonically varied.
- Real part of the frequency dependent complex heat capacity is related to the freezing-in of an order parameter characterizing a peculiar internal degree of freedom of the sample. This effect depends on the ratio of the kinetic relaxation time constant of the degree of freedom as compared to the time scale of the perturbation.
- Imaginary part of the frequency dependent complex heat capacity has no particular physical meaning. Nevertheless, the entropy produced during the irreversible process, averaged over the time scale of the measurement, is directly proportional to the imaginary part of the complex impedance of the measurement, which is the imaginary part of the inverse of the complex measured heat capacity. Also, the imaginary part of the complex impedance is equal to zero at zero-frequency (reversible experiment) and equal to zero at infinite frequency (irreversibility maximum). The imaginary part of the complex heat capacity has the same behavior, and we can conclude that it may also be a representation of heat dissipation or heat lost during the experiment.

By analogy with heat dissipation during thermal diffusion processes, where heat is absorbed along a spatial axis, we claim that during irreversible calorimetric experiments, a certain amount of heat is lost along the path over a peculiar virtual axis represented by the internal order parameter (degree of advance of the reaction in the peculiar case of chemical reactions) representing a certain internal degree of freedom (the advance of the reaction in the peculiar case of chemical reactions). This view can be applied to any irreversible experiments. For example in dielectric relaxations, the internal parameter may be the angle between the electric field and the polarization vector response.

The generality of this previous approach has been demonstrated. Although thermodynamics of irreversible processes due to chemical reactions has been first considered by De Donder, it can concern all physico-chemical transformations or relaxation phenomena occurring out-of equilibrium (first order phase transition, glass transition, relaxation phenomena…) that are induced by temperature and characterized by a state variable (internal order parameter) characterizing a certain internal degree of freedom of a sample. On a general manner, this approach has also been applied to dynamic DSC experiments, assuming that thermodynamic internal thermal equilibrium is reached. In this case, a beginning of explanation of the experimentally measured heat capacity during glass transitions has been envisaged. A special focus has been done on the link existing between imaginary part of the inverse of the complex heat capacity and the finite amount of entropy produced during non-equilibrium temperature modulated heat capacity measurements. The notion of entropy produced during one period of the oscillation in temperature modulated calorimetric experiments has been clarified.

In summary, we can conclude that the notion of frequency dependent complex heat capacity must be very useful in ac-calorimetry experiments for the study of lots of type of transitions and thermal phenomena.

This work was realized inside the Groupe de Biothermique et de Nanocalorimétrie of the CRTBT. The author wants to thank O. Bourgeois, G. Gaudin, J. Richard, H. Guillou, for useful discussions and many corrections of this paper, Z. Atkinson and C. Garden for English corrections and A. Michetti for bibliographic researches. I am infinitely indebted to J. Chaussy for having passed on me his passion and knowledge of theoretical and experimental calorimetry.

**Symbols**

**LATIN**

| | |
|---|---|
| $a$ | angular coefficient of De Donder |
| $A$ | chemical affinity |
| $A_0$ | constant affinity |
| A, B,… | thermodynamic states of a system; initial and final products in a reaction sequence |
| $C^*$ | complex heat capacity |
| $C'$ | real part of the complex heat capacity |
| $C''$ | imaginary part of the complex heat capacity |
| $C_\infty$ | contribution to the heat capacity of the infinitely fast degree of freedom |
| $C_0$ | contribution to the heat capacity at equilibrium of all the degree of freedom |
| $C_p$ | heat capacity at constant pressure |
| $C_{mes}$ | experimentally measured heat capacity |
| $C_\xi$ | heat capacity at constant composition or at constant order parameter |
| $C_{rev}$ | experimentally measured heat capacity during a reversible calorimetric experiment |
| $d_eS$ | infinitesimal external entropy exchange |
| $d_iS$ | infinitesimal internal entropy creation |
| $G$ | free energy of Gibbs |
| $H$ | enthalpy or heat content function |
| $K$ | heat exchange coefficient |
| $L$ | phenomenological coefficient of Onsager |
| $N$ | number of mole of a constituent |
| $P$ | heat flow rate or thermal power; pressure |
| $P_i$ | thermal power of irreversibility or rate of the uncompensated heat of Clausius |

$\overline{P}_i$ time averaged thermal power of irreversibility

$P_0$ amplitude of the oscillating thermal power

$Q$ heat

$Q'$ uncompensated heat of Clausius

$R$ constant of perfect gas

$S$ entropy

$t$ time

$T$ temperature; period of the modulation

$T_{ac}$ oscillating temperature

$T_{dc}$ constant stationary temperature

$v$ rate of reaction

**GREEK**

$\gamma$ temperature rate

$\delta A$ amplitude of the oscillating affinity

$\delta T_{ac}$ amplitude of the oscillating temperature

$\Delta\xi$ departure from equilibrium of the variable $\xi$ after the time interval $\Delta t$

$\Delta\xi_{eq}$ distance between two equilibrium values of the variable $\xi$ after the time interval $\Delta t$

$\eta_{eq}$ contribution to the measured heat capacity at equilibrium of an internal degree of freedom

$\kappa$ phase lag generated by irreversible effects on the oscillating temperature

$\mu$ chemical potential

$\nu$ stochiometric coefficient

$\xi$ degree of advance of a reaction or order parameter of an internal degree of freedom

$\xi_{eq}$ equilibrium value of the degree of advance of a reaction or equilibrium value of an order parameter of an internal degree of freedom

$\sigma_i$ instantaneous rate of production of entropy

$\overline{\sigma}_i$ time averaged rate of production of entropy

$\tau$ kinetic relaxation time constant of an internal degree of freedom

$\tau_{ext}$ kinetic relaxation time constant of the temperature towards the heat bath

$\tau_{int}$ kinetic relaxation time constant of the temperature inside a medium due to thermal diffusion

$\varphi$ phase of the oscillating temperature

$\phi$ phase of the oscillating affinity

$\omega$ angular frequency of the oscillating thermal power or oscillating temperature

Fig. 1: A thermodynamic system, in thermal and mechanical equilibrium, with a heat capacity *C*, and a well-defined temperature *T*, is linked to a thermal bath at a constant temperature $T_0$ via a heat loss coefficient *K*. A known quantity of heat *Q* is supplied to the system or received from the system by the experimentalist.

Fig. 2: The temporal evolution of the thermodynamic system is represented by a curve in the diagram ($T, \xi$). After a perturbation generated by the experimentalist (via $\delta Q$), the system is driven from a thermodynamic state A to another thermodynamic state B. Along this thermodynamic pathway, the amount of heat exchanged between the system and the outside world is linked to one contribution of the entropy variation of the system ($d_eS = \delta Q/T$). This entropy exchanged over the boundaries of the system can be positive or negative. At the same time, a quantity of heat (the uncompensated heat of Clausius) is produced inside the system. This heat is linked to the other contribution of the entropy variation of the system ($d_iS = \delta Q'/T$). It is due to irreversible processes occurring within the system and it is always positive. It is equal to zero along the reversible pathway.

Fig. 3: This figure is a Fresnel's diagram in which three time dependent oscillating vectors are represented. The x axis is given by the phase of the oscillating input thermal power taken by convention as the phase reference ($\varphi = 0$). The y axis is given by the phase $\varphi = \pi/2$. $\varphi$ is the phase of the oscillating temperature. The first vector is the oscillating temperature with its two components (their values are provided) projected on the x and y axis. The second vector is the vector time derivative of the oscillating temperature with a phase advance of $\pi/2$. The firth vector is the oscillating affinity with its two components (their values are provided) projected on the new axis represented by the two preceding vectors.

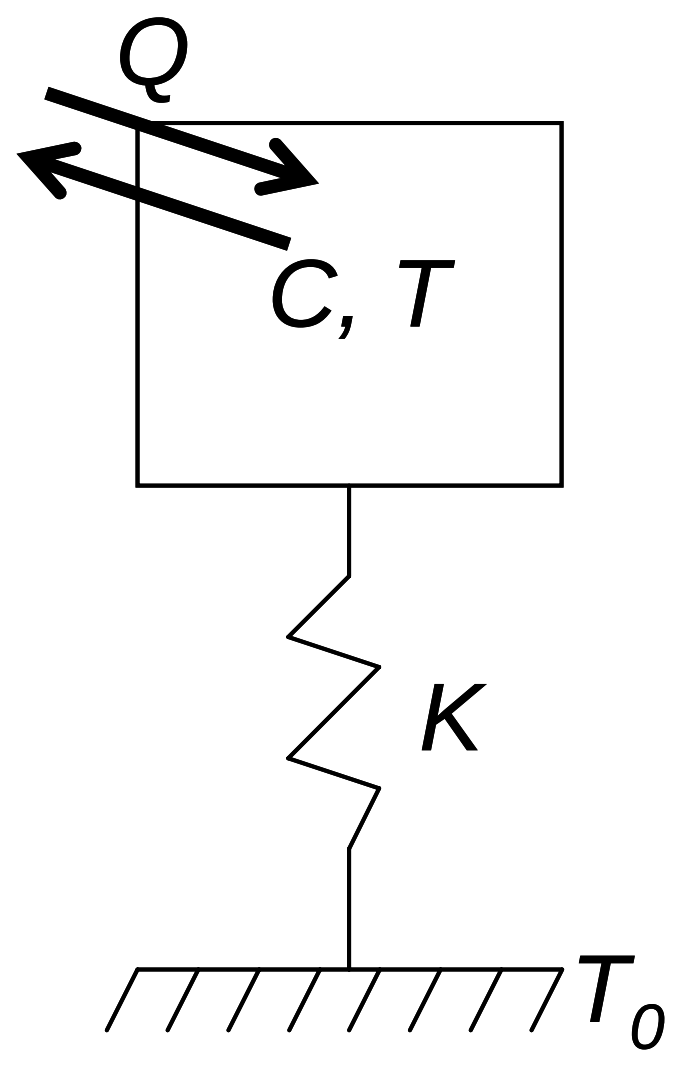

J.-L. Garden, Thermochimica Acta Fig1

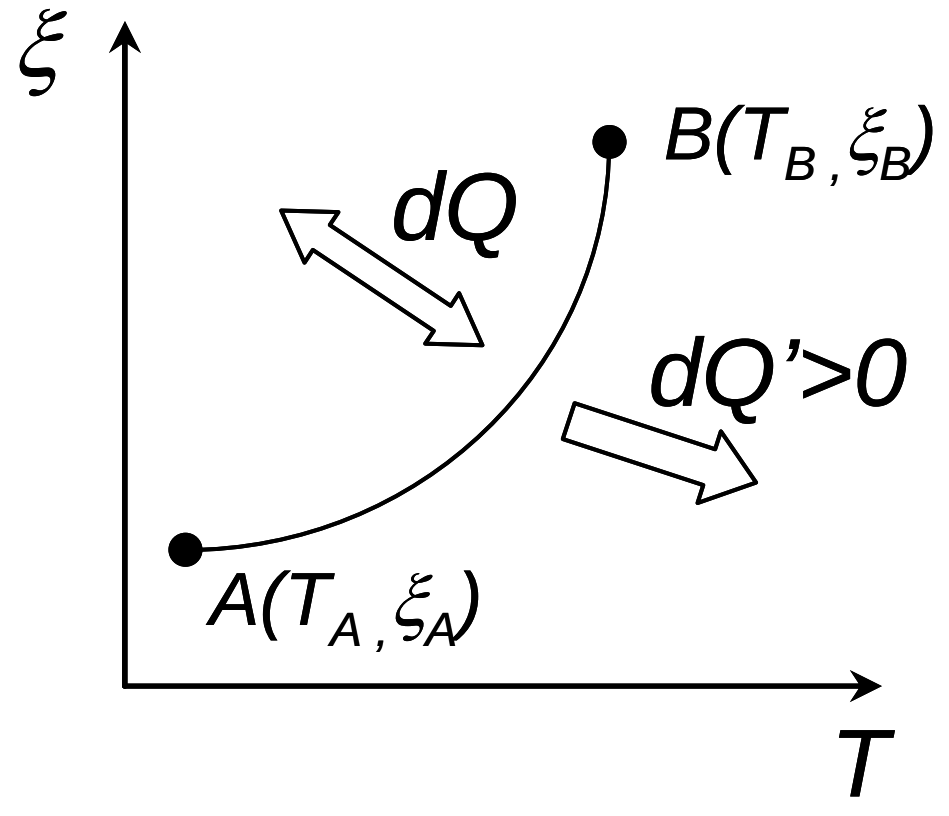

J.-L. Garden, Thermochimica Acta Fig2

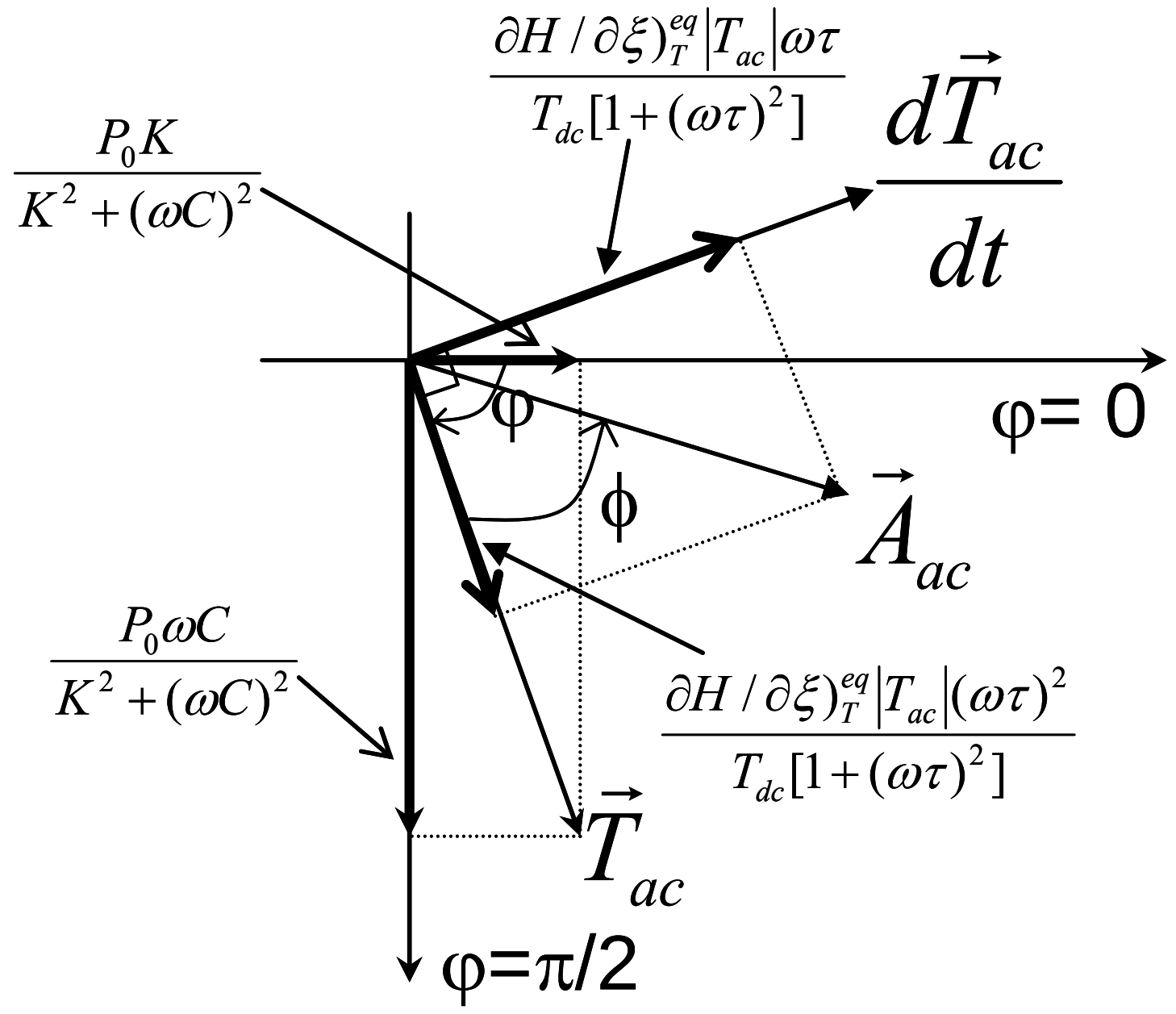

J.-L. Garden, Thermochimica Acta Fig 3